\documentclass{PoS}
\rightline{\parbox{4cm}{\large\rm
DESY 09-196\\
Edinburgh 2009/20\\
LTH 853\\
TUM-T39-11-11
}}

\usepackage{amsmath}
\usepackage{amssymb}
\usepackage{graphicx}
\usepackage{wrapfig}
\newcommand{\be}{\begin{equation}}
\newcommand{\ee}{\end{equation}}
\newcommand{\bc}{\begin{center}}
\newcommand{\ec}{\end{center}}

\title{Lattice Investigations of Nucleon Structure at Light Quark Masses}

\ShortTitle{Nucleon Structure at Light Quark Masses}

\author{M.~G\"ockeler$^{a}$, Ph.~H\"agler$^{b}$, R.~Horsley$^{c}$, Y.~Nakamura$^{a}$,
  D.~Pleiter$^{d}$, P.~E.~L.~Rakow$^{e}$, A.~Sch\"afer$^{a}$, G.~Schierholz$^{a,f}$,
  H.~St\"uben$^{g}$, \speaker{J.~M.~Zanotti}$^{c}$\\
        \llap{$^a$} Institut f\"ur Theoretische Physik,
                    Universit\"at Regensburg,
                    93040 Regensburg, Germany \\
        \llap{$^b$} Institut f\"ur Theoretische Physik T39,
                    TU M\"unchen,
                    85747 Garching, Germany \\
        \llap{$^c$} School of Physics and Astronomy,
                    University of Edinburgh,
                    Edinburgh EH9 3JZ, UK \\
        \llap{$^d$} John von Neumann Institute NIC / DESY Zeuthen,
                    15738 Zeuthen, Germany \\
        \llap{$^e$} Theoretical Physics Division,
                    Department of Mathematical Sciences,
                    University of Liverpool,
                    Liverpool L69 3BX, UK \\
        \llap{$^f$} Deutsches Elektronen-Synchrotron DESY,
                    22603 Hamburg, Germany \\
        \llap{$^g$} Konrad-Zuse-Zentrum f\"ur Informationstechnik Berlin,
                    14195 Berlin, Germany \\
        E-mail: \email{jzanotti@ph.ed.ac.uk}}

\author{QCDSF/UKQCD Collaboration}

      \abstract{Lattice simulations of hadronic structure are now
        reaching a level where they are able to not only complement,
        but also provide guidance to current and forthcoming
        experimental programmes at, e.g. Jefferson Lab, COMPASS/CERN
        and FAIR/GSI. 
        By considering new simulations at low quark masses and on
        large volumes, we review the recent progress that has been
        made in this exciting area by the QCDSF/UKQCD collaboration.
        In particular, results obtained close to the physical point
        for several quantities, including electromagnetic form factors and
        moments of ordinary parton distribution functions, show some
        indication of approaching their phenomenological values.
      }

\FullConference{The XXVII International Symposium on Lattice Field Theory - LAT2009\\
                 July 26-31 2009\\
                 Peking University, Beijing, China}

\begin{document}

%
\section{Introduction}
\label{sec:intro}
%

Much of our knowledge about hadronic structure in terms of quark and
gluon degrees of freedom has been obtained from high energy scattering
experiments.
However, there are still many unresolved issues in
hadronic physics that need to be addressed, from both an experimental
and theoretical perspective.
This is one of the main motivations of the 12~GeV Jefferson Lab
upgrade which aims to \cite{JLab}: search for exotic mesons;
study the role of hidden flavours in the nucleon; map out the spin and
flavour dependence of the valence parton distribution functions;
explore nuclear medium effecs; and measure the generalised parton
distribution functions of the nucleon. 
It is imperative that these and other exciting experimental efforts,
such as those at COMPASS/CERN and FAIR/GSI, are matched by modern
lattice simulations.
See recent lattice reviews \cite{Orginos:2006zz, Hagler:2007hu,
  Zanotti:2008zm, Renner:2009} for latest developments.

\begin{figure}
\bc
\vspace*{-3mm}
\includegraphics[width=0.75\textwidth]{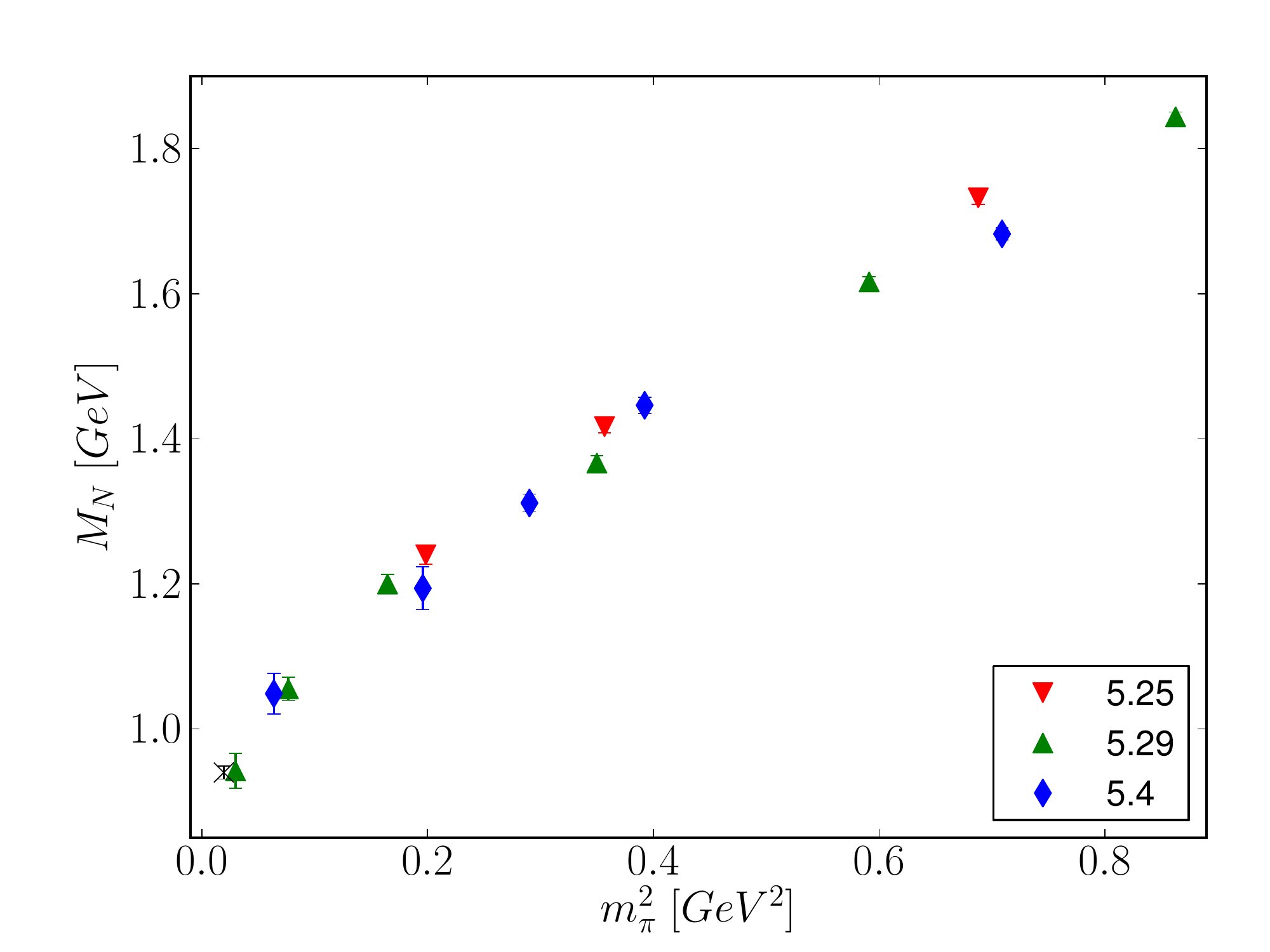}
\caption{Nucleon mass as a function of $m_\pi^2$ for three different
  values of $\beta$. The scale is set using
  $r_0=0.467$~fm and the physical mass is indicated by the cross.}
\label{fig:mn}
\vspace*{-3mm}
\ec
\end{figure}
In this talk we present the latest results from the QCDSF/UKQCD
Collaboration for electromagnetic (EM) form factors, $\langle
x\rangle_q,\ \langle x\rangle_{\Delta q},\ g_A$ and $g_T$.
Simulations are performed with the Wilson gauge action and two
flavours of ${\cal O}(a)$-improved Wilson fermions.
Four values of $\beta=5.20,\,5.25\,,5.29\,,5.40$, corresponding to
lattice spacings in the range $0.1<a<0.07$~fm, allow for the approach
to the continuum limit to be assessed, while a range of lattice
volumes ($1.1<L<3.2$~fm) enable us to search for finite size effects
in our simulations.
A key feature of the new results presented here is the control we are
now able to achieve over the approach to the chiral limit.
This is made possible by several new simulations with pion masses
reaching as low as 170~MeV, which are summarised in
Fig.~\ref{fig:mn}.

%
\section{Electromagnetic Form Factors}
\label{sec:ff}
%

The study of the electromagnetic properties of hadrons provides
important insights into the non-perturbative structure of QCD.
The EM form factors reveal important information on the internal
structure of hadrons including their size, charge distribution and
magnetisation.

A lattice calculation of the $q^2$-dependence of hadronic
electromagnetic form factors can not only allow for a comparison with
experiment, but also help in the understanding of the asymptotic
behaviour of these form factors, which is predicted from perturbative
QCD.
Such a lattice calculation would also allow for the extraction of
other phenomenologically interesting quantities such as charge radii
and magnetic moments.
For a recent review see \cite{Arrington:2006zm}.

On the lattice, we determine the form factors $F_1(q^2)$ and
$F_2(q^2)$ by calculating the following matrix element of the
electromagnetic current
\be
\langle p',\,s'| j^{\mu}(\vec{q})|p,\,s\rangle
\, = 
 \bar{u}(p',\,s')
 \left[ \gamma^\mu F_1(q^2) +
       i\sigma^{\mu\nu}\frac{q_\nu}{2M_N}F_2(q^2) \right] 
 u(p,\,s) \, ,
\label{eq:em-me}
\ee
where $u(p,\,s)$ is a Dirac spinor with momentum, $p$, and spin
polarisation, $s$, $q = p' - p$ is the momentum transfer, $M_N$ is
the nucleon mass and $j_\mu$ is the electromagnetic current.
The Dirac $(F_1)$ and Pauli $(F_2)$ form factors of the proton are
obtained by using $j_\mu^{(p)} = \frac{2}{3}\bar{u}\gamma_\mu u -
\frac{1}{3}\bar{d}\gamma_\mu d$, while for isovector form factors
$j_\mu^v = \bar{u}\gamma_\mu u - \bar{d}\gamma_\mu d$.
It is common to rewrite the form factors $F_1$ and $F_2$ in terms of
the 
electric and magnetic Sachs form factors, 
$G_e= F_1 + q^2/(2M_N)^2\, F_2$ and $G_m= F_1 + F_2$.

If one is using a conserved current, then (e.g. for the proton)
$F_1^{(p)}(0) = G_e^{(p)}(0) =1$ gives the electric charge,
while $G_m^{(p)}(0) = \mu^{(p)} = 1 + \kappa^{(p)}$
gives the magnetic moment, where $F_2^{(p)}(0) = \kappa^{(p)}$ is the
anomalous magnetic moment.
From Eq.~(\ref{eq:em-me}) with see that $F_2$ always appears with a
factor of $q$, so it is not possible to extract a value for $F_2$ at
$q^2=0$ directly from our lattice simulations.
Hence we are required to extrapolate the results we obtain at finite
$q^2$ to $q^2=0$.
Form factor radii, $r_i=\sqrt{\langle r_i^2\rangle}$, are defined from
the slope of the form factor at $q^2=0$.

\begin{figure}[t]
     \vspace*{-7mm}
     \hspace{-0.2mm}
   \begin{minipage}{0.49\textwidth}
      \centering
          \includegraphics[clip=true,width=1.05\textwidth]{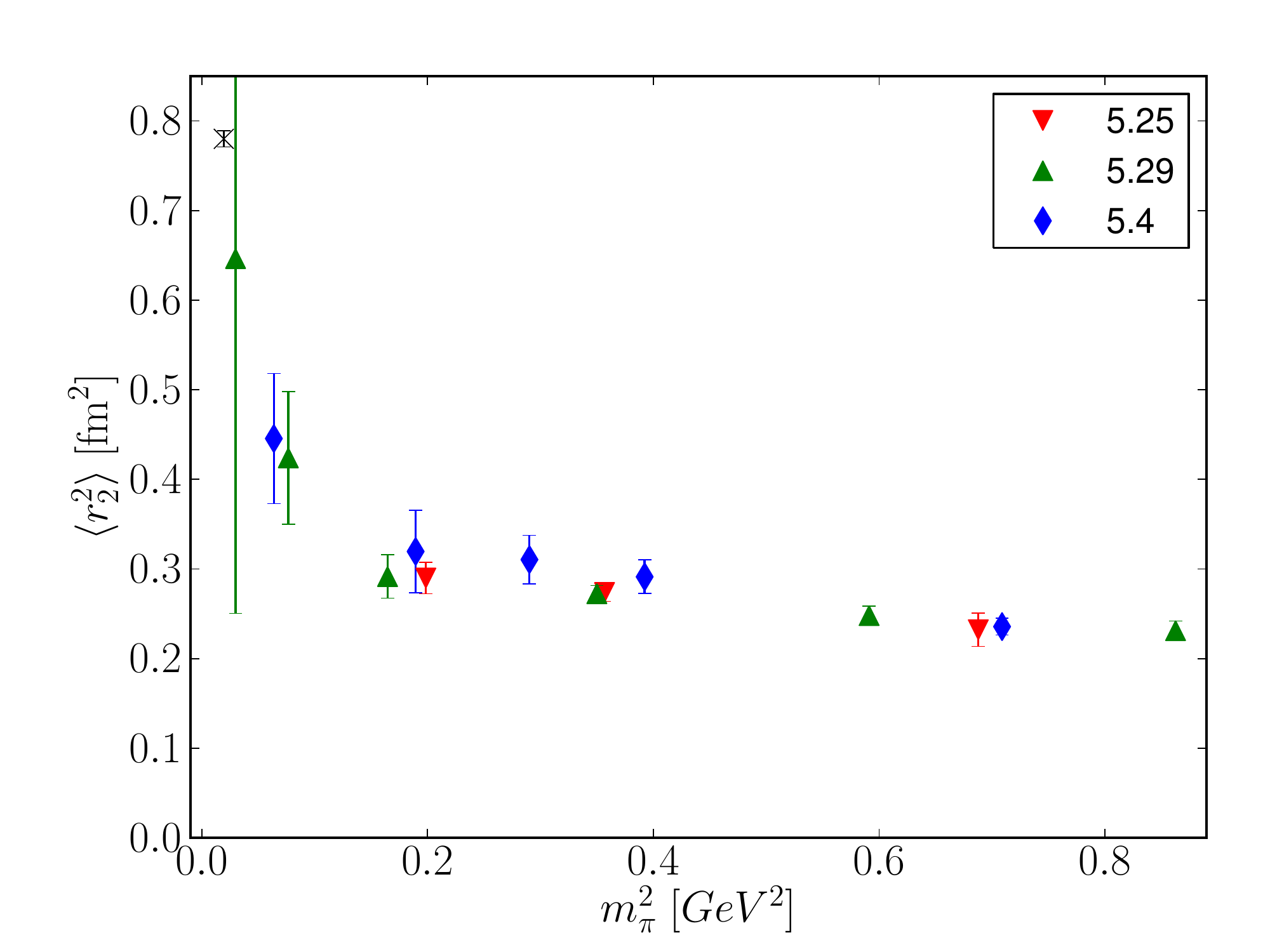}
\caption{Results for the isovector Pauli radius, $r_2$.}
\label{fig:r2v}
     \end{minipage}
    \begin{minipage}{0.49\textwidth}
      \centering
          \includegraphics[clip=true,width=1.05\textwidth]{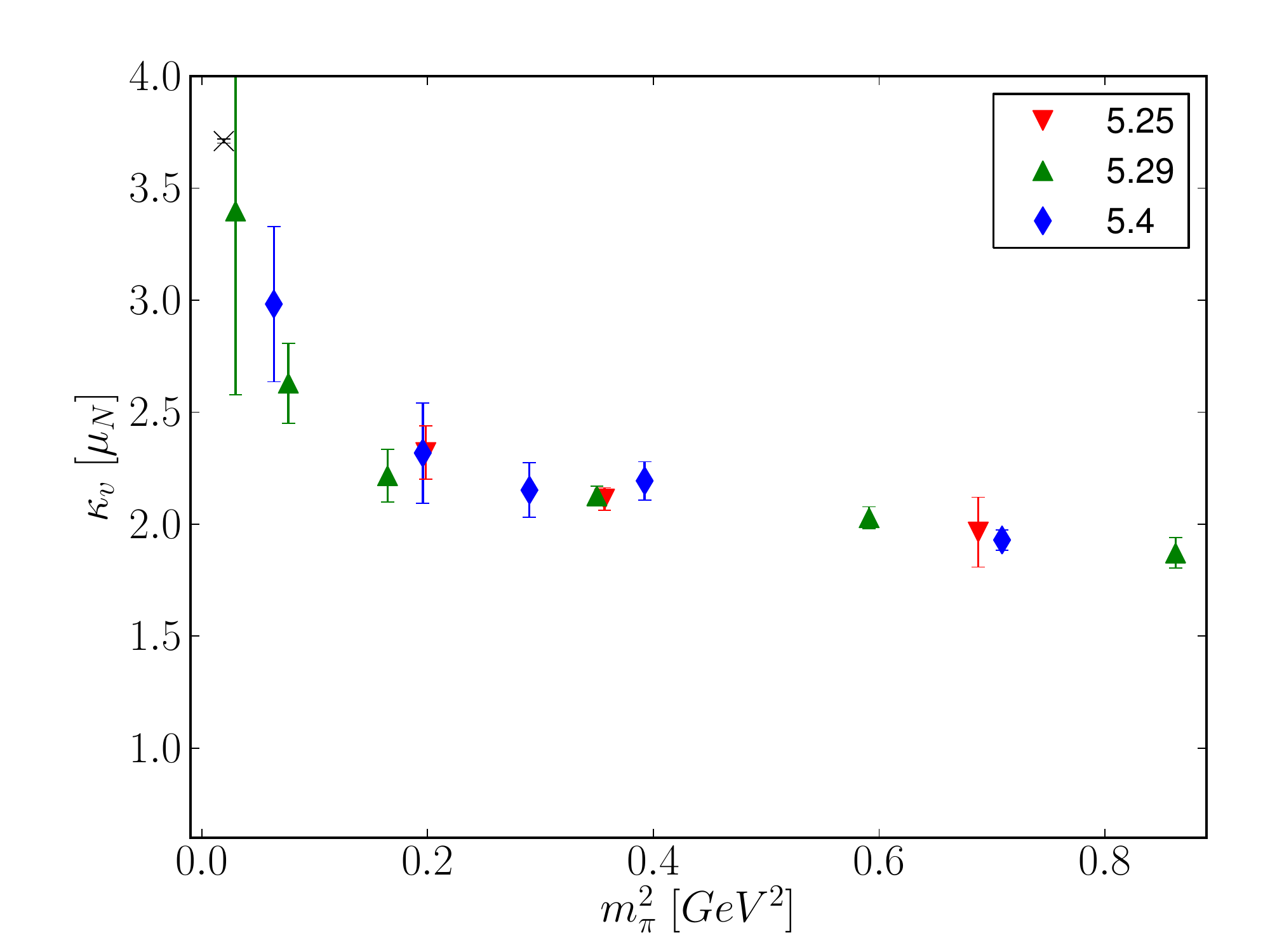}
\caption{Isovector anomalous magnetic moment, $\kappa_v$ in nuclear magnetons.}
\label{fig:kv}
     \end{minipage}
 \end{figure}

In Fig.~\ref{fig:r2v} we see results for the isovector Pauli
radius plotted as a function of $m_\pi^2$.
Here we observe a common trend of lattice determinations of this quantity,
namely that the results are a factor of $\sim 3$ smaller than
experiment and show little variation as a function of $m_\pi^2$ for pion masses
larger than $m_\pi>300$~MeV.
While the new results below $m_\pi<300$~MeV are showing signs of
upward curvature, as predicted from chiral perturbation theory
\cite{Gockeler:2003ay,Young:2004tb,Hemmert:2009}, it is not yet clear whether this
is enough to reach the experimental value.
Of course, at such light quark masses, finite volume effects are
expected to suppress the nucleon charge radii \cite{Hemmert:2009}, and
the results from three different volumes at
$m_\pi^2\approx0.09$~GeV$^2$ indicate that this is indeed the case,
although the effect appears to be rather small at this pion mass.

Results for the isovector anomalous magnetic moment, $\kappa_v$, are
shown in Fig.~\ref{fig:kv} and the results are seen to be slowly
increasing towards the experimental value as the pion mass is
decreased.
Closer to the chiral limit, the results are starting to exhibit stronger
chiral behaviour, once again as predicted by $\chi$PT
\cite{Gockeler:2003ay,Young:2004tb,Hemmert:2009}.
Here finite volume effects are also expected to suppress the magnetic
moment \cite{Young:2004tb}, so the fact that our results on a finite volume
are not increasing fast enough to agree perfectly with experiment is
not totally unexpected.

%
\section{Moments of Structure Functions}
\label{sec:SF}
%

In this section we discuss our latest findings for the nucleon's axial
charge, $g_A$, tensor charge, $g_T$, and spin-independent and
spin-dependent quark momentum fractions, $\langle x\rangle_q$ and
$\langle x\rangle_{\Delta q}$.
All operators in this section have been renormalised nonperturbatively
using the Rome-Southampton method \cite{NPR}.

\subsection{Axial Charge, $g_A$}
\label{sec:gA}

The axial coupling constant of the nucleon is important as it governs
neutron $\beta$-decay and also provides a quantitative measure of
spontaneous chiral symmetry breaking.
It is also related to the first moment of the helicity dependent quark
distribution functions, $g_A=\Delta u - \Delta d$.
It has been studied theoretically as well as experimentally for many
years and its value, $g_A=1.2695(29)$, is known to very high accuracy.
Hence it is an important quantity to study on the lattice, and since
it is relatively clean to calculate (zero momentum, isovector), it
serves as useful yardstick for lattice simulations of nucleon
structure.

The axial charge is defined as the value of the isovector axial form
factor at zero momentum transfer and is determined by the
forward matrix element
\be 
\langle p,\,s|A_{\mu}^{u-d}|p,\,s\rangle \, = 2g_A s_\mu \ ,
\ee
where $A_\mu=\bar q \gamma_\mu \gamma_5 q$, $p$ is the nucleon momentum,
and $s_\mu$ is a spin vector with $s^2=-M_N^2$.

\begin{figure}[t]
     \vspace*{-7mm}
    \begin{minipage}{0.48\textwidth}
     \hspace*{-7mm}
      \centering
          \includegraphics[clip=true,width=1.12\textwidth]{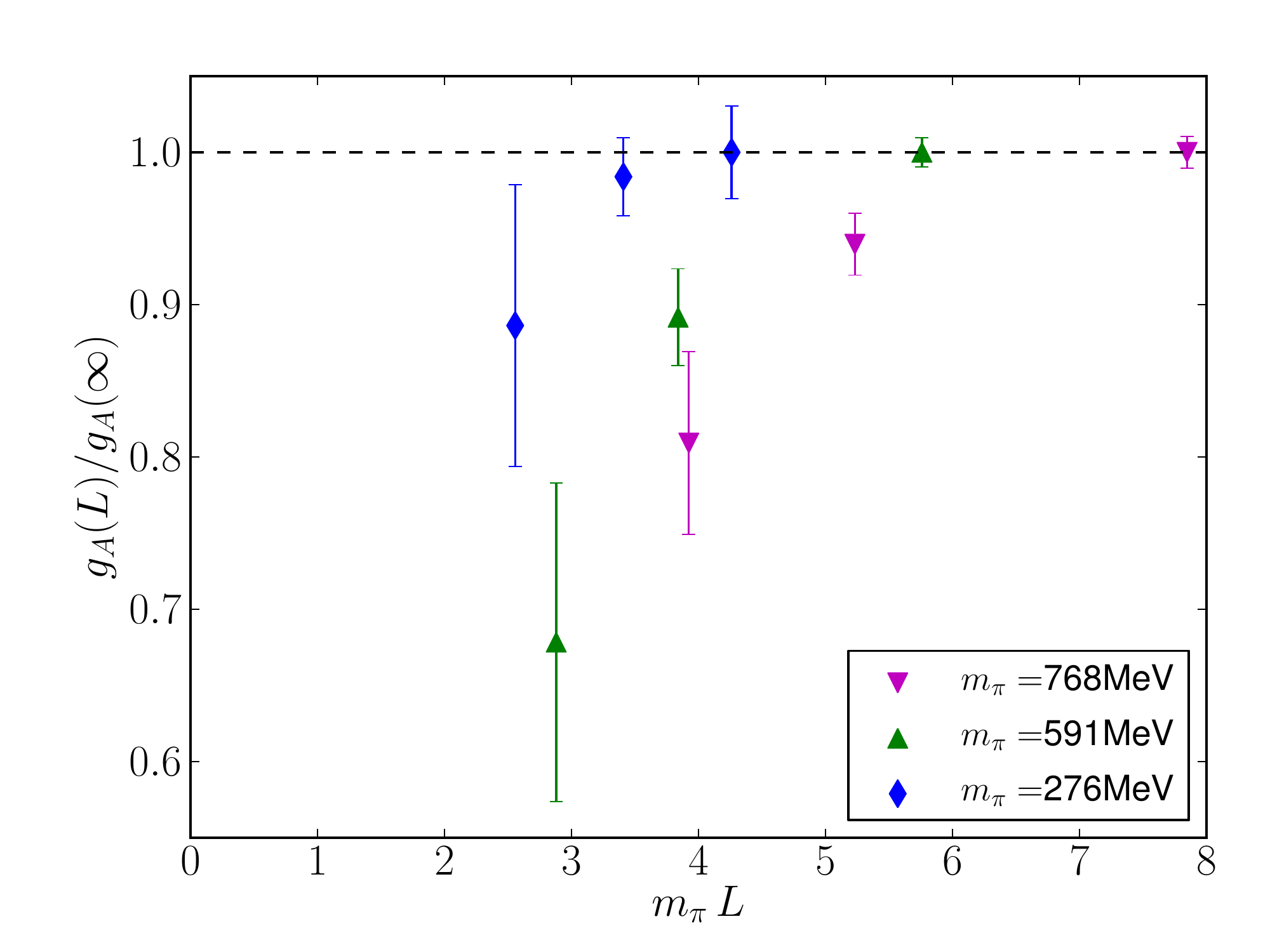}
          \caption{Finite volume results for the nucleon axial charge,
            $g_A(L)$, normalised with the largest available volume,
            $g_A(\infty)$, plotted as a function of $m_\pi L$ for
            $\beta=5.29$.}
\label{fig:gAmpiL}
     \end{minipage}
     \hspace{4mm}
   \begin{minipage}{0.48\textwidth}
     \vspace*{-2mm}
     \hspace*{-6mm}
      \centering
          \includegraphics[clip=true,width=1.12\textwidth]{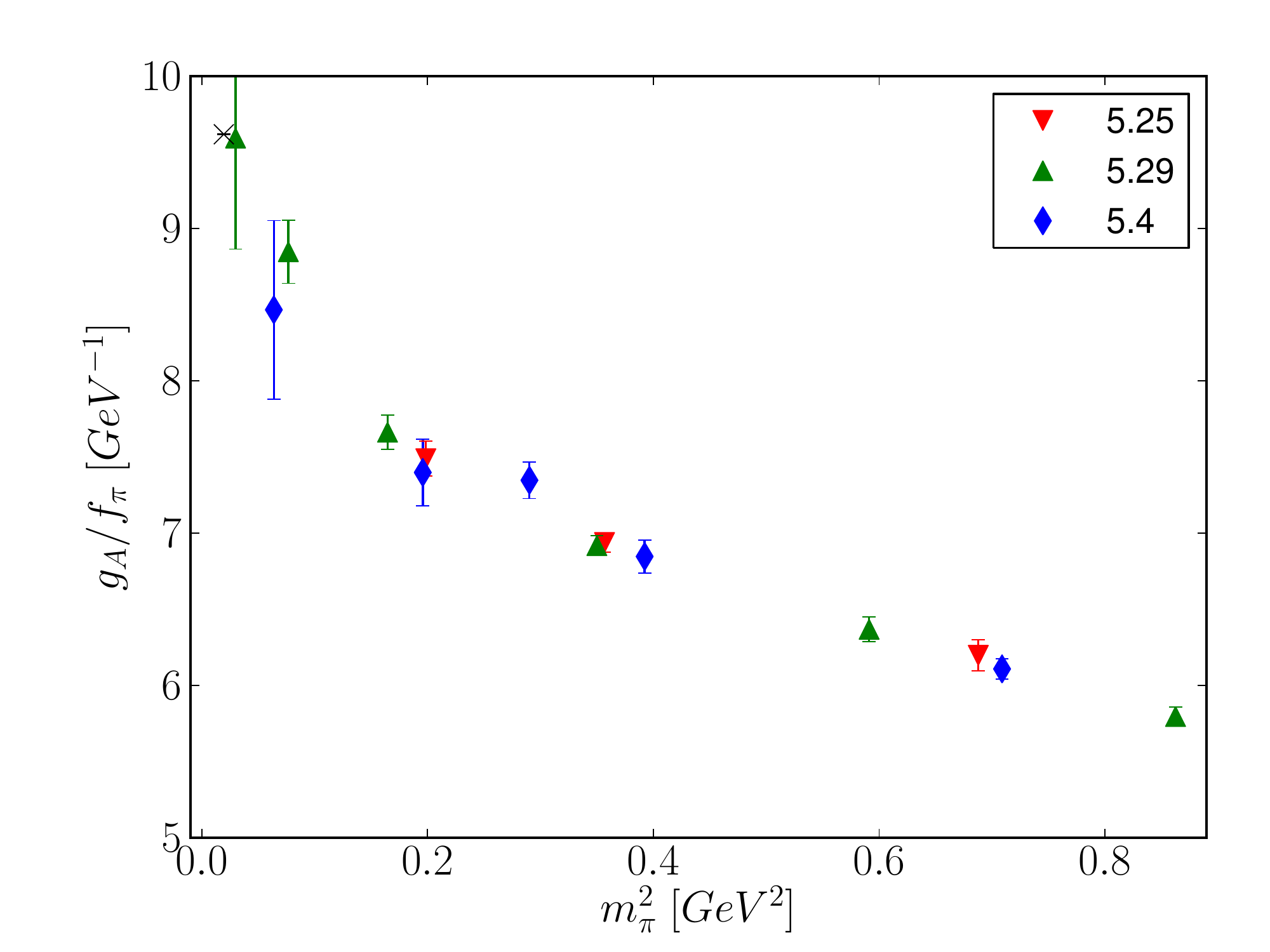}
\caption{Results for the ratio of the nucleon axial charge and the
  light pseudoscalar decay constant, $g_A/f_\pi$}
\label{fig:gA}
     \end{minipage}
 \end{figure}

 $g_A$ has been studied in-depth for many years by the QCDSF/UKQCD
 \cite{Khan:2006de}, LHP \cite{Edwards:2005ym} and RBC/UKQCD
 \cite{Yamazaki:2008py} collaborations and has been shown to suffer
 from large finite size effects.
 Such effects are also seen in our new results at lighter pion masses
 as seen in Fig.~\ref{fig:gAmpiL} where we show those results for
 where we have more than one volume available at a fixed pion mass.
 Here we normalise the finite volume results for the axial charge,
 $g_A(L)$, with the result obtained from the largest available volume
 for that choice of $(\beta,\kappa)$, $g_A(\infty)$.
%

Another way to visualise our results is to consider the ratio
$g_A/f_\pi$, as shown in Fig.~\ref{fig:gA}.
A good reason for considering this ratio is that the common
renormalisation constant, $Z_A$, cancels in the ratio.
It is also a natural ratio to consider in the context of the
Goldberger-Treiman relation, $g_A=f_\pi g_{\pi NN}/M_N$, which is
valid in the chiral limit and approximately valid at larger quark
masses.
As can be seen in Fig.~\ref{fig:gA}, the results for $g_A/f_\pi$ show
a strong dependence on $m_\pi^2$ and increase towards the experimental
value as they approach the physical point. 
This may be an indication that the leading finite size effects cancel
in this ratio \cite{Khan:2006de,Colangelo:2005gd}.
It would be nice to see if such cancellations occur in $\chi$PT
expressions of these quantities.
This is currently under investigation.

\subsection{Tensor Charge, $g_T$}
\label{sec:gT}

Although the tensor charge has not received as much attention as $g_A$,
there have been recently some new attempts to calculate the quantity
on the lattice \cite{Gockeler:2005cj,Lin:2008uz,Ohta:2008kd}.
Since $g_T$ is not known experimentally, it provides an opportunity
for the lattice to make a prediction, although given the difficulties with
$g_A$ as discussed above, care must be taken, not only with the chiral
extrapolation, but also in the assessment of finite size effects.

In Fig.~\ref{fig:gT} we show the latest QCDSF/UKQCD results, including the
new simulations at low pion masses.
Here we observe a similar trend to that observed in the axial charge,
$g_A$, above, namely that increasing the volume at a fixed pion mass
increases the result, although the effects are not as strong.

Considering the behaviour of the results at fixed $m_\pi L$, we
observe that the results are reasonably flat as a function of
$m_\pi^2$.
This is to be compared with the above result for $g_A$ where we
observed a definite increase in the results.

Given that we still don't have the chiral and finite volume
systematics under control, it is not yet possible to make a strong
prediction for the isovector tensor charge of the nucleon, however our
current results indicate that it is likely to be in the range
$0.9<g_T<1.1$.

\subsection{Nucleon Momentum Fraction, $\langle x\rangle$}
\label{sec:x}

Lattice studies of $\langle x\rangle_{q}$ are 
notorious in
that all lattice results to date at heavy quark masses exhibit an
almost constant behaviour in quark mass towards the chiral limit and
are almost a factor of 1.5 larger than phenomenologically accepted
results, e.g. $\langle x\rangle_{u-d}^{\rm MRST}= 0.157(9)$
\cite{Martin:2001es}.
Despite predictions that the results should drop dramatically in the
light quark mass region \cite{Detmold:2001jb}, we are still yet to see
conclusive evidence for this from a lattice calculation.

To date, only connected contributions have been simulated to high
precision, and although progress is being made towards the evaluation
of disconnected contributions \cite{Bali}, here we only quote results
for isovector quantities where disconnected contributions cancel.

\begin{figure}[t]
     \vspace*{-7mm}
    \begin{minipage}{0.49\textwidth}
      \centering
          \includegraphics[clip=true,width=1.05\textwidth]{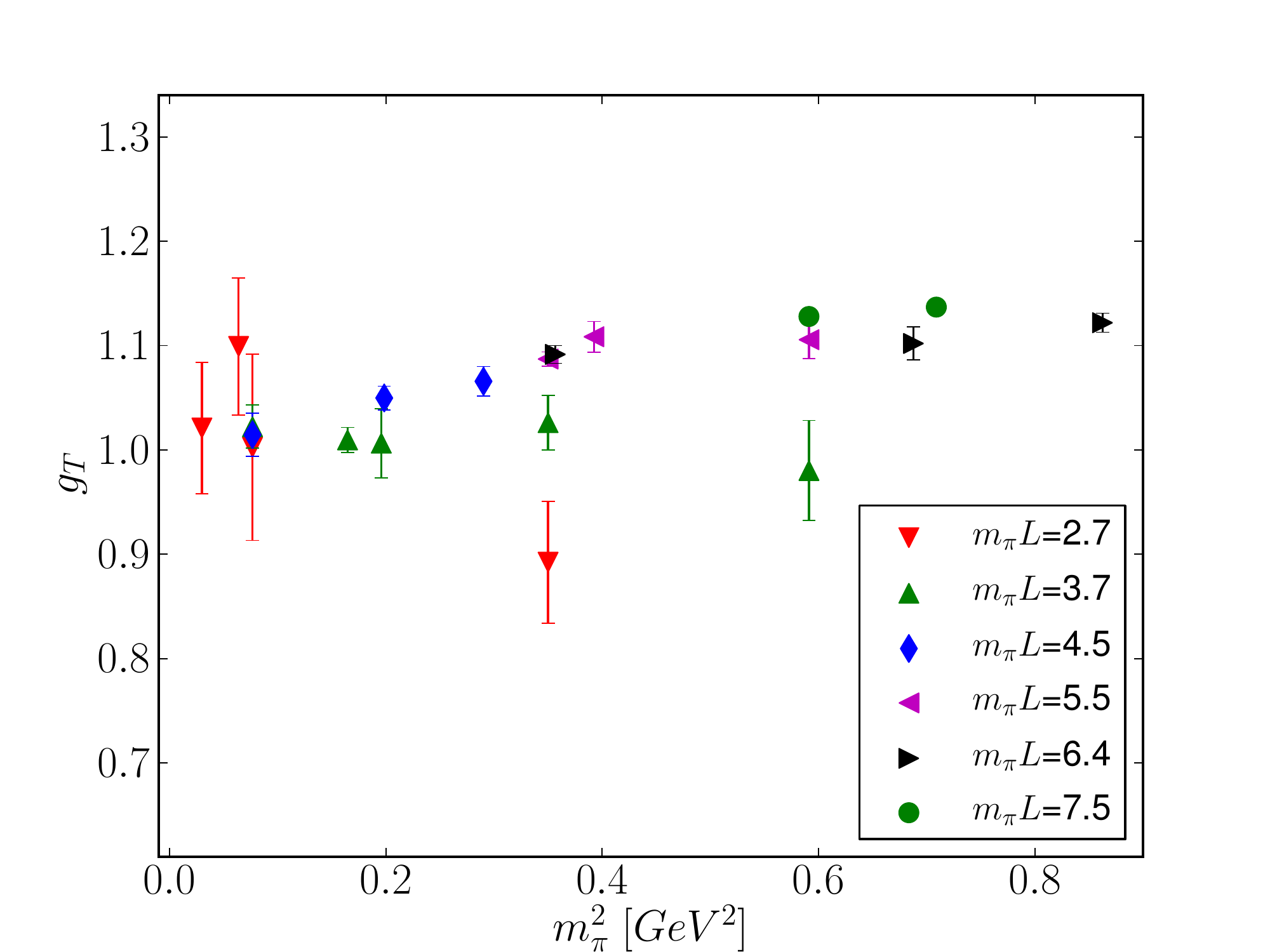}
\caption{Results for the isovector tensor charge, $g_T$, in the $\overline{\rm MS}$ scheme at $\mu=2$ GeV.}
\label{fig:gT}
     \end{minipage}
   \begin{minipage}{0.49\textwidth}
      \centering
          \includegraphics[clip=true,width=1.05\textwidth]{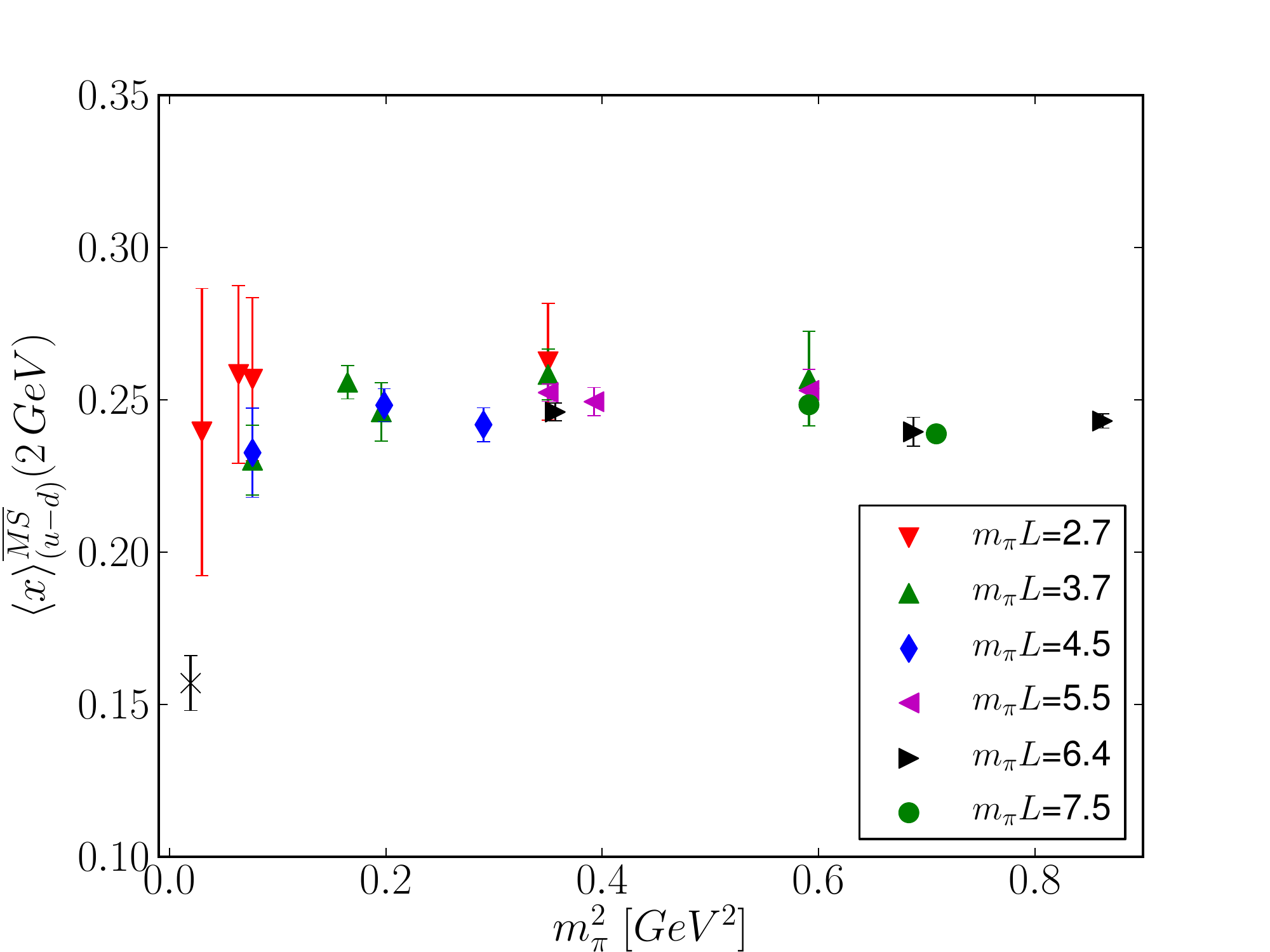}
\caption{Results for the isovector quark momentum fraction, $\langle
  x\rangle_{u-d}$, in the $\overline{\rm MS}$ scheme at $\mu=2$ GeV.}
\label{fig:xq}
     \end{minipage}
 \end{figure}
Figure~\ref{fig:xq} shows the latest QCDSF/UKQCD results including new
results below $m_\pi<300$~MeV.
While the results above $m_\pi^2>0.1$~GeV$^2$ show the same constant
behaviour seen in many lattice simulations, the new results in the
region below $m_\pi^2<0.1$~GeV$^2$ are beginning to show some
interesting behaviour, albeit with larger statistical errors.
What is particularly interesting is the results at a fixed pion mass
($\approx 270$~MeV), but with three different lattice volumes
($L=24,\,32,\,40$).
Here we observe that the result from the smallest volume continues the
trend of the results at heavier pion masses, while the results from
the larger volumes sit slightly lower.
This is not totally unexpected, as the dramatic downward curvature has
been shown to be only achievable on larger volumes
\cite{Hemmert:2009,Detmold:2003rq,Detmold:2005pt}.
For this reason, we expect that even after a large increase in
statistics, the current point at $m_\pi\approx170$~MeV with $L=40$
could well still lie somewhat higher than phenomenological
determinations.
A further simulation at this pion mass but with spatial volume length
of $L=64$ is currently underway to further test these ideas.

Similar behaviour is seen in the spin-dependent momentum fraction,
$\langle x\rangle_{\Delta q}$.

%
\section{Conclusion \& Outlook}
\label{sec:fin}
%

We have presented some of the latest results on nucleon structure from
the QCDSF/UKQCD collaboration, including EM form factors, $\langle
x\rangle_q,\, \langle x\rangle_{\Delta q},\,g_A$ and $g_T$.
These calculations are now becoming available at pion masses as low as
$m_\pi\approx 170$\,MeV, so direct comparison with experimental
determinations will soon be possible.
However, as we have seen in, e.g. $g_A$, finite size effects are
starting to become a serious issue.
As a result, we are now planning a new simulation on a volume of
$(5\,{\rm fm})^3$, in order to minimise these effects, although
corrections from $\chi$PT will still probably need to be taken into
account.

%
\vspace*{-3mm}
\section*{Acknowledgements}
%
\vspace*{-3mm}

The numerical calculations have been performed on the APE1000 and
apeNEXT at NIC/DESY (Zeuthen, Germany), the IBM BlueGeneL at EPCC
(Edinburgh, UK), the BlueGeneL and P at NIC (J\"ulich, Germany) and
the SGI Altix 4700 at LRZ (Munich, Germany). 
Some of the configurations at the small pion mass have been generated
on the BlueGene/L at KEK by the Kanazawa group as part of the DIK
research programme. We thank all institutions. 
This work has been supported in part by the EU Integrated
Infrastructure Initiative Hadron Physics (I3HP) under contract
RII3-CT-2004-506078 and by the DFG under contracts FOR 465
(Forschergruppe Gitter-Hadronen-Ph\"anomenologie) and SFB/TR 55
(Hadron Physics from Lattice QCD).  
PH is supported by the DFG under the Emmy-Noether program. 
JZ is supported through the UK's {\it STFC Advanced Fellowship
  Programme} under contract number ST/F009658/1.

%

\end{document}